\documentclass[%
 reprint,
 amsmath,amssymb,
 aps,
floatfix,
]{revtex4-2}

\usepackage{graphicx}
\usepackage{dcolumn}
\usepackage{bm}

\begin{document}

\preprint{APS/123-QED}

\title{Dissipative switching waves and solitons in the systems with spontaneously broken symmetry}

\author{D. Dolinina}
\email{d.dolinina@metalab.ifmo.ru}
\author{A. Yulin}%
\email{a.v.yulin@corp.ifmo.ru}
\affiliation{ITMO University, Saint-Petersburg 197101, Russia
}%

\date{\today}

\begin{abstract}
The paper addresses the bistability caused by spontaneous symmetry breaking bifurcation in a one-dimensional periodically corrugated  nonlinear waveguide pumped by coherent light at normal incidence. The formation and the stability of the switching waves connecting the states of different symmetries are studied numerically. It is shown that the switching waves can form stable resting and moving bound states (dissipative solitons). The protocols of the creation of discussed nonlinear localized waves are suggested and verified by numerical simulations. 
\end{abstract}

\maketitle


\section{Introduction}

It has been known that in bistable nonlinear systems two different spatially uniform states can be connected by a switching wave  preserving its shape. These structures have first been reported in  \cite{first_fronts} and then found in many optical systems \cite{pattern_review, polarization_walls,walls_staliunas}. The integrity of the domain wall is supported by exact balance of different linear and nonlinear effects such as diffraction, the dependency of the effective refractive index on the light intensity, linear and nonlinear losses, the external pump, etc. In these respect the domain walls are similar to nonlinear localized waves called solitons. It is important to note that the domain walls connecting different spatially uniform states can be at rest (this is so-called Maxwell point) but the general case is when domain walls are moving. The motion of a domain wall results in the expansion of one of the uniform states and to the shrinking of another and this is why these localized structures got the name of switching waves. Let as remark that these switching waves are also often referred as domain walls because they separate different states. In the present paper we use both terms. 

If two or more domain walls have formed in a system then they can interact to each other provided that the distance between the domain walls is comparable to their characteristic size. In some cases this interaction can result in the formation of bound states of the switching waves. These bound states can also be seen as dissipative solitons \cite{rosanov_autosolitons, bound_st6, rosanov_review, bound_st1, bound_st2, bound_st3, bound_st4, bound_st5}. The dissipative solitons play an important role in optics, for example they are used for the generation of femtosecond pulses in fiber lasers \cite{femto_pulses1,femto_pulses2}. This explains why dissipative optical solitons have been actively studied for many years. 

In the present paper we consider dissipative localized waves in optical systems where there may exist modes of different symmetries that have very different losses. These systems are related to so-called optical bound states in continuum (BIC) that are actively studied now because this phenomenon opens a way to achieve solitary high-Q resonances. BIC systems have been already used for second and third harmonic generation \cite{harmonic_gen1, harmonic_gen2} and in laser design \cite{lasing_BIC}.

The characteristic feature of BIC states is that they cannot be directly excited by the external coherent light, therefore they are often referred as dark states (DS). The states coupled to free propagating waves have higher losses but can be directly excited, these states are called bright states (BS). It is obvious that in the presence of the finite intrinsic losses no stationary states can have a structure of pure DS. However, the BS can be unstable against the linear excitations having a structure of DS. This instability breaks the symmetry of the solution leading to the formation of hybrid states (HS) that can be seen as a combination of BS add DS. It is important to note that the dominating component of HS can be of DS kind and, thus, HS can experience very low effective losses and so they can exist at lower levels of pump intensities. This can facilitate the experimental observation of optical bistability and other nonlinear effects.

For our purposes we do not need a true BIC when for some modes the radiative losses vanish completely but a quasi-BIC when the radiative losses of some modes are significantly depressed by destructive interference. In general case of quasi-BIC the quasi-DS can weakly interact with non-guided waves but within the model used in this paper these losses are accounted as effective intrinsic losses of DS. Recently it was shown that bright dissipative solitons can nestle on the HS in the systems of such a kind, see \cite{OL_BIC_solitons}. The aim of the present paper is to study the domain walls connecting different spatially uniform states. It will be shown that the domain walls connecting HS and BS exist, that they can be dynamically stable and that they can form stable bound states.

It is well known that the symmetric systems can have asymmetric soliton solutions and that the symmetry breaking set the dissipative solitons in motion  \cite{walk_weiss, walk_fedorov, walk_scroggie, moving_sols_rosanov1,moving_sols_rosanov3, walk_turaev, walk_tlidi, walk_Egorov, walk_cortes}. Let us remark that symmetry breaking bifurcation is also known for connecting two physically equivalent spatially uniform states. In this case the symmetry breaking transform a resting Ising-kind domain wall into a moving domain wall of Bloch kind, \cite{ising_coullet, ising_michaelis, ising_staliunas, ising_gomila}. In our paper we discuss both the resting and moving dissipative solitons but in our case the motion of the solitons is caused by the broken symmetry of the solitons pedestals and the type of the soliton pedestal defines the direction of the soliton motion.

The article is organized as follows. In the next section we briefly discuss the physical system and formulate the mathematical model describing it. In this section we also summarize important facts on the structure and the stability of the spatially uniform states. The formation and the stability of the switching waves connecting spatially uniform states of different symmetries are studied in the third section. Fourth section is devoted to the bound states of the switching waves, i.e. to dissipative solitons. In this section we also suggest protocols that allow to observe the dissipative solitons in the experiments.  Finally, in the conclusion we list the main results of the paper. 

\section{The physical system and its mathematical model}

As a system possessing quasi-BIC states we consider externally pumped one-dimensional waveguides with periodical grating and Kerr nonlinearity. The resonant scattering on the periodical grating results in the appearance of a gap in the dispersion characteristics and to dramatic decrease of the eigenwaves group velocity which becomes equal to zero at the exact resonance. It is worth noting, that the upper and the lower modes can experience very different radiative losses. A simple explanation of this is that each mode can be seen as a composition of two counter-propagating waves. Each of the counter-propagating waves can leak from the waveguide but the total radiative losses are defined by the interference of the contributions from each of the counter-propagating waves. In the case of destructive interference the contributions from the counter-propagating waves cancel each other and, thus, the radiative losses get suppressed. The constructive interference enhances the radiative losses.

The dynamics of the system is described as in \cite{Krasikov_BIC} by two counter-propagating waves approach and can be expresses mathematically by following system of equations: 
\begin{eqnarray}
(\partial_t \pm \partial_x) U_{\pm}=&&(i\delta-\gamma) U_{\pm} + i\alpha (|U_{\pm}|^2 +2|U_{\mp}|^2)U_{\pm} + \nonumber\\&&+ (i\sigma - \Gamma)(U_{\pm}+U_{\mp}) +P,
\label{main_1}
\end{eqnarray}

where $U_{+}$ and $U_-$ are the slow varying complex amplitudes of the two counter-propagating waves, $\delta$ is the detuning of the frequency of the pump from the centre of the gap of the dispersion characteristics, $\gamma$ is the losses of different nature that cannot be compensated by the $U_{+}$ and $U_{-}$ waves interference, $\alpha=\pm 1$ is the Kerr-nonlinearity coefficient, $\sigma$ is the conservative part of the coupling coefficient defining the rate of the mutual re-scattering of the counter-propagating waves, $\Gamma$ is the dissipative coupling accounting the fact that the radiative losses depend on the interference of the waves.

It is convenient to reformulate the problem in terms of the complex amplitude of a bright $U_b=\frac{U_{+}+U_{-}}{\sqrt{2}}$ and a dark modes $U_d=\frac{U_{+}-U_{-}}{\sqrt{2}}$. In this variables equation (\ref{main_1}) reads:
\begin{eqnarray}
\label{bright}
(\partial_t -i\delta -i\dfrac{3}{2}K+ \gamma +2\Gamma - 2 i\sigma) U_b +(\partial_x  + &&iM)U_d= \nonumber\\=&&2P
\end{eqnarray}
\begin{eqnarray}
(\partial_t -i\delta -i\dfrac{3}{2}K+ \gamma) U_d +(\partial_x  +iM)U_b=0,
\label{dark}
\end{eqnarray}
where $K=\alpha(|U_b|^2+|U_d|^2)$, $M=\alpha Re (U_b U_d^{*})$. 
From (\ref{bright},\ref{dark}) it is seen that linear eigenmodes in the form $e^{-i \omega t + i k x}$ with $k = 0$ have the structure ($U_b \neq 0$, $U_d =0$) and ($U_b = 0$, $U_d \neq 0$). The second mode has only intrinsic losses $\gamma_d=\gamma$ and, thus, the losses are lower than the losses of the first mode $\gamma_b=\gamma+2\Gamma$. It is also obvious from the equation (\ref{dark}) that the second mode cannot be excited by the driving force accounting for the action of pumping wave at normal incidence. Thus, this mode is the dark one. The first mode can be excited by the pump and so it is a bright mode. In the nonlinear regime there may exist the states such that both their components are non-zero, $U_b \neq 0$, $U_d \neq 0$, \cite{Krasikov_BIC,OL_BIC_solitons}. These modes are the hybrid nonlinear modes.   

\begin{figure}[b]
\centering
\fbox{\includegraphics[width=\linewidth]{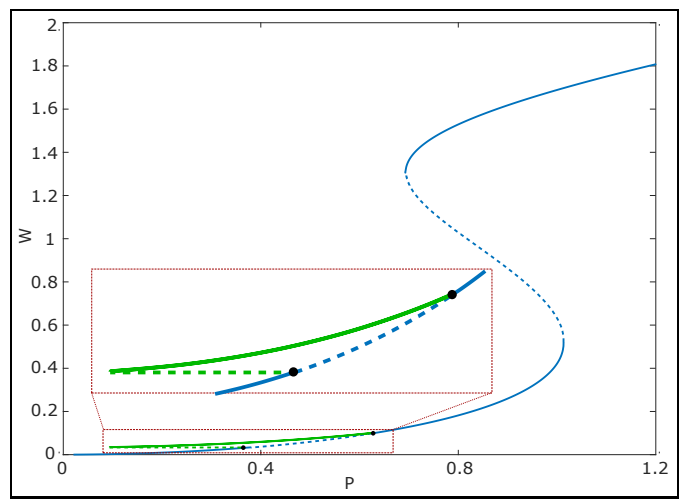}}
\caption{ Bifurcation diagram of the bright (blue line) and hybrid (green line) states, where $W = |U_b|^2 + |U_d|^2 = |U_+|^2 + |U_-|^2$ is intensity of states; dynamically unstable solutions are shown by dashed line. Parameters are: $\delta = 0.05$, $\alpha = -1$, $\sigma = 1$, $\gamma = 0.001$, $\Gamma = 0.299$.}
\label{fig1}
\end{figure}

As it is shown in \cite{Krasikov_BIC} the spatially uniform hybrid states are always unstable for $\alpha = 1$ and, therefore, we consider only the case $\alpha=-1$. To facilitate the discussion of the nonlinear localized waves and, in particular, switching waves we briefly reproduce the results on the formation and stability of the spatially uniform states reported in \cite{OL_BIC_solitons}.  Stationary nonlinear states can be classified as bright states (BS) with $U_b \neq 0$, $U_d=0$ and as hybrid states (HS) with $U_{b,d} \neq 0$. The bifurcation diagram of the BS is shown in Fig.~\ref{fig1} by the blue curve. The HS bifurcating from the BS are shown by the green curve. The spectral linear stability analysis as well as direct numerical simulations show that the HS belonging to the upper branch of the bifurcation curve are stable and, thus, are of interest from physical point of view.  

Here we acknowledge an important fact that the hybrid states are produced by the spontaneous symmetry breaking and, therefore, they consist of two counter-propagating waves of different amplitudes $|U_{+}| \neq |U_{-}|$ and so these are the states with energy flow $F_{HS}=|U_{+}|^2-|U_{-}|^2$ directed either from the left to the right or from the right to the left. The symmetry of the equations $x \rightarrow -x$, $U_{\pm} \rightarrow U_{\mp}$ insures that if $U_{+}=a$, $U_{-}=b$ is a spatially uniform solution then $U_{+}=b$, $U_{-}=a$ is also a spatially uniform solution. Thus, the spatially uniform HS are double degenerate and have the energy flux of the same absolute value but of different signs. 


\section{Domain walls connecting the hybrid and bright states }

In this Section we consider the stationary domain walls connecting different spatially uniform HS to spatially uniform BS. Let us start with introducing the notation of the domain walls. In the present paper we focus on the domain walls connecting the stable hybrid states to the bright states belonging to the lower branch of the bifurcation characteristics. Then it is possible to mark the domain walls by the kinds of the spatially uniform states connected by the domain walls. For example a domain wall ``$BH$'' is a connection of the bright state on the left to the hybrid state on the right. Then ``$HB$'' is a domain wall where the bright and the hybrid states are swept (HS is on the left side of the domain wall and BS is on the right). It is important to note that the characteristic size of the domain walls should be much larger than the grating period, otherwise slowly varying amplitude approximation is not applicable and new effects appear \cite{rosanov_chan}.

As it is mentioned above, the hybrid states have non-zero energy flux and so we need to distinguish the hybrid states where energy flows from the left to the right from the hybrid states with opposite direction of the energy flow. The former ones we denote as $H_{+}$ and the latter as $H_{-}$. The bright states have zero energy flux and will be marked as $B$ without indices. Let us emphasise, that from physical point of view it is obvious that the domain walls $H_{+}B$ and $H_{-}B$ are different, in the first case the energy flow in the HS is directed towards the domain wall, in the second case - away from the domain wall.

\begin{figure}[t]
\centering
\fbox{\includegraphics[width=\linewidth]{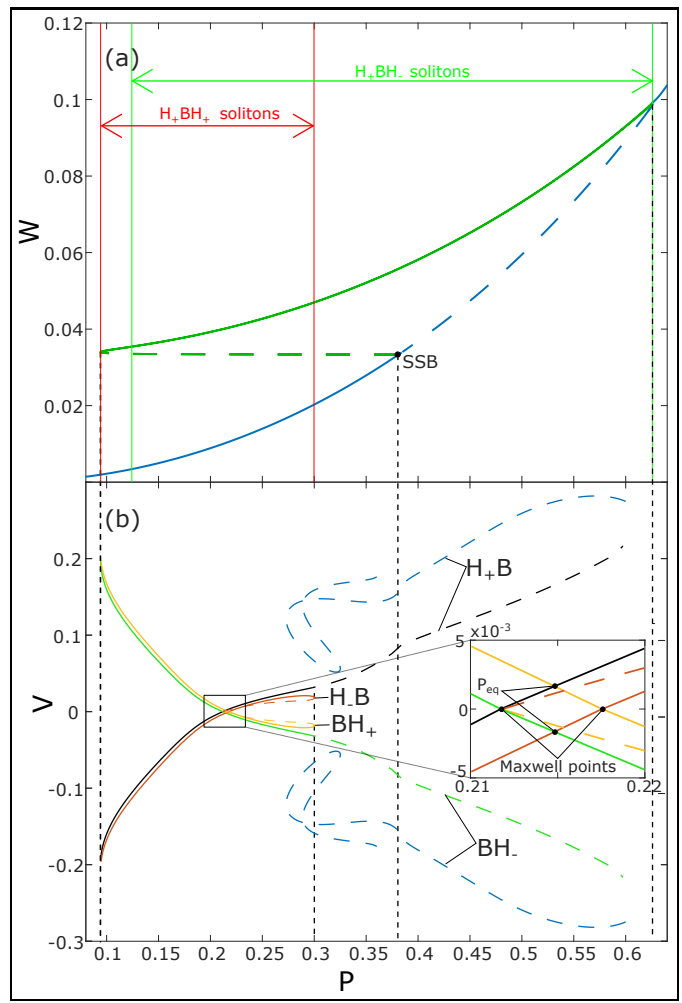}}
\caption{(a) The part of the bifurcation diagram showing the dependencies of the intensities of the hybrid and bright (belonging to the lower branch) spatially uniform states on the pump amplitudes. Two vertical dashed lines mark the range of the existence of the stable hybrid states. The green lines and arrows show the range of the existence of resting $H_{+}BH_{-}$ and $H_{-}BH_{+}$ solitons. The range of existence of moving $H_{+}BH_{+}$ and $H_{-}BH_{-}$ solitons is shown by the red lines and arrows.  The point where the bright states becomes unstable is marked as "SSB". (b) Velocity dependencies of the domain walls on pump. The bifurcation curves of the different domain walls are shown by different colors. The solid lines correspond to the stable and the dashed lines to the unstable domain walls. The inset shows the enlarged region near the Maxwell points of $H_+B$ and $BH_+$ domain walls. The points where the domain walls $H_+B$ and $BH_+$ (or $H_-B$ and $BH_-$) have equal velocities are marked as `$P_{eq}$'.}
\label{fig2}
\end{figure}

The linear analysis of the stationary perturbations on the H$_\pm$S and BS background has revealed that the small excitations can exponentially decay with or without oscillations to the backgrounds. Therefore one can expect that the states can be connected by domain walls with exponential tails. As it is mentioned before, the domain walls move with some velocity which depends on the parameters of the system. The assumption that the domain wall is moving without changing its shape allows to look for this solution in a moving reference frame where the domain wall field distribution depends only on the coordinate $\xi=x-vt$ and so is described by ordinary differential equations. The equations have to be solved with the boundary conditions corresponding to the prescribed spatially uniform states on the left and on the right. From the mathematical point of view a domain wall is a heteroclinic phase trajectory connecting two different equilibrium points in the phase space of the differential equation written in the moving reference frame. These trajectories do not exist for an arbitrary velocity but for some values of $v$ the heteroclynic trajectories can be found. We numerically found the domain walls connecting the upper hybrid state to the lower bright state; a typical bifurcation diagram $v$ vs $P$ of the domain walls is shown in panel (b) of Fig.~\ref{fig2}. 

Let us first discuss the domain walls $H_{+}B$ connecting the state $H_{+}$ on the left and $B$ on the right. The dependency of the velocity of the domain walls on the pump is shown for typical parameters in  Fig.~\ref{fig2}(b) by the black line. One can see that there is a special value of the pump called a Maxwell point (due to analogy with thermodynamics)  at which the velocity of the domain wall is zero. The existence of resting domain walls are known in many physical systems, including optical ones \cite{first_fronts, pattern_review, polarization_walls, yulin_fronts}. The existence of Maxwell point in our system is important for the formation of the dissipative solitons (bound states of domain walls) that are considered in the next Section in detail. 

\begin{figure}[t]
\centering
\fbox{\includegraphics[width=\linewidth]{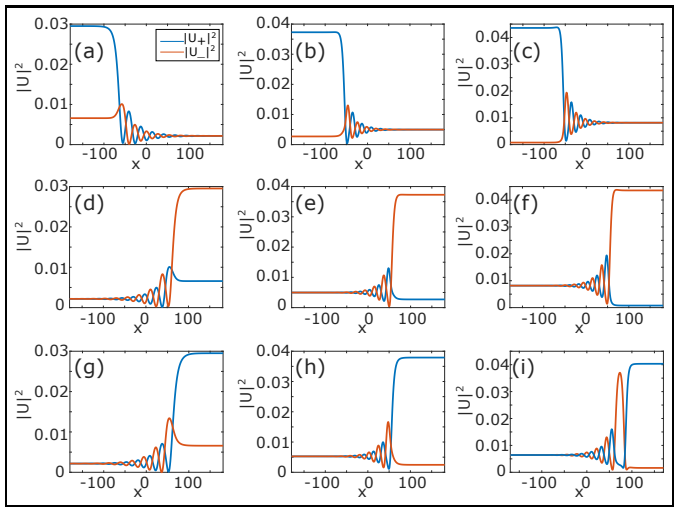}}
\caption{Field distributions of the domain walls. (a) $H_+B$ at $P = 0.14$; (b) $H_+B$ at Maxwell point; (c) $H_+B$ at $P = 0.27$; (d) $BH_-$ at $P = 0.14$; (e) $BH_-$ at Maxwell point; (f) $BH_-$ at $P = 0.27$; (g) $BH_+$ at $P = 0.14$; (h) $BH_+$ at Maxwell point; (i) $BH_+$ at $P = 0.24$ of the unstable branch.}
\label{fig3}
\end{figure}

The range of the existence of $H_{+}B$ almost coincides with the range of the existence of the stable HS (the upper branch of the HS bifurcation curve shown in Fig.~\ref{fig2}(a)). The domain wall is wide for the low pump intensities, numerical simulations indicate that the domain wall width goes to infinity at the left border of existence domain. Typical distributions of the domain wall fields are shown in Fig.~\ref{fig3}(a)-(c) for different intensities of the pump.

Let us remark here that it is also possible to find other family of $H_{+}B$ domain walls, see Fig.~\ref{fig2}(b) where the corresponding branch of the bifurcation diagram is shown by black dashed line. These domain walls have more complex structure but we did not manage to find stable domain walls of such a kind.

Another important remark is that because of the symmetry $x \rightarrow -x$, $U_{\pm} \rightarrow U_{\mp}$ of the equations (\ref{main_1}) the domain walls $BH_{-}$ can be obtained from the domain walls $H_{+}B$ by the inversion of the direction of $x$ axis and the swap of the fields $U_{\pm}$, compare panels (a)-(c) and (d)-(f) of Fig.~\ref{fig3} showing the field distributions in $H_{+}B$ and $BH_{-}$ domain walls for the same pump intensities. For a fixed pump the velocities of $H_{+}B$ and $BH_{-}$ domain walls are of the same absolute value but of different sign, see the bifurcation diagrams of the domain walls presented in Fig.~\ref{fig2}(b)

\begin{figure}[b]
\centering
\fbox{\includegraphics[width=\linewidth]{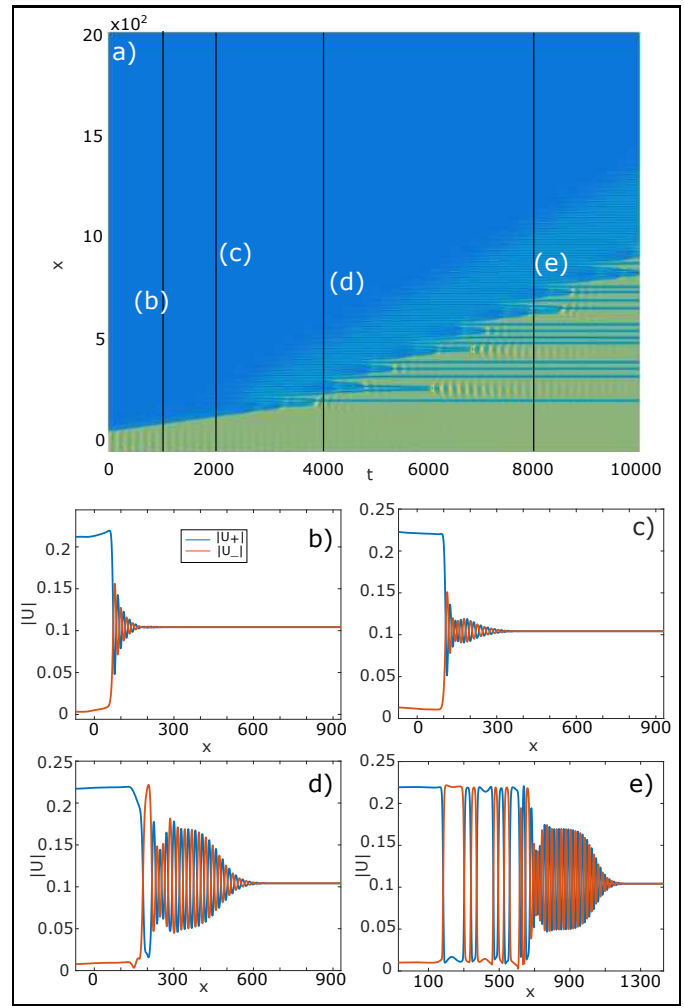}}
\caption{(a) The evolution of the unstable $H_+B$ wall at $P = 0.31$ obtained by numerical simulations of equations (\ref{main_1}). The colors shows the total intensity of the state $|U_{+}|^2+|U_{-}|^2$, the blue color corresponds to the lowest intensity and the yellow color - to the highest intensity. (b) - (e) The distribution of the fields at different times.}
\label{fig4}
\end{figure}

Now let us discuss an important issue of the dynamical stability of the domain walls. It is obvious that if one of the backgrounds of the domain wall is unstable then the whole state is unstable and, thus, the domain walls existing to the right from the point of SSB (Spontaneous Symmetry Breaking bifurcation destabilizing the lower bright state) are always unstable. However, the linear stability analysis shows that the domain wall loses its stability at the pump values lower then that of SSB. For relatively low pump intensity the backgrounds remain stable but the domain wall gets destroyed, see Fig.~\ref{fig4} showing the development of the instability. 

The instability results in the development of oscillating pattern separating the HS and BS. At longer times the hybrid state expands but stable dark solitons stays incorporated in the HS. The interaction between the dark soliton leads to their collision and annihilation, this process is seen well in Fig.~\ref{fig4}. However, the interaction between the solitons decreases exponentially with the distance. Therefore, the HS with incorporated solitons can be seen as a  metastable state. We remark here that the oscillation pattern can be an indication of the existence of a periodic nonlinear state but this problem requires a separate consideration and is out of the scope of the present paper.

\begin{figure}[t]
\centering
\fbox{\includegraphics[width=\linewidth]{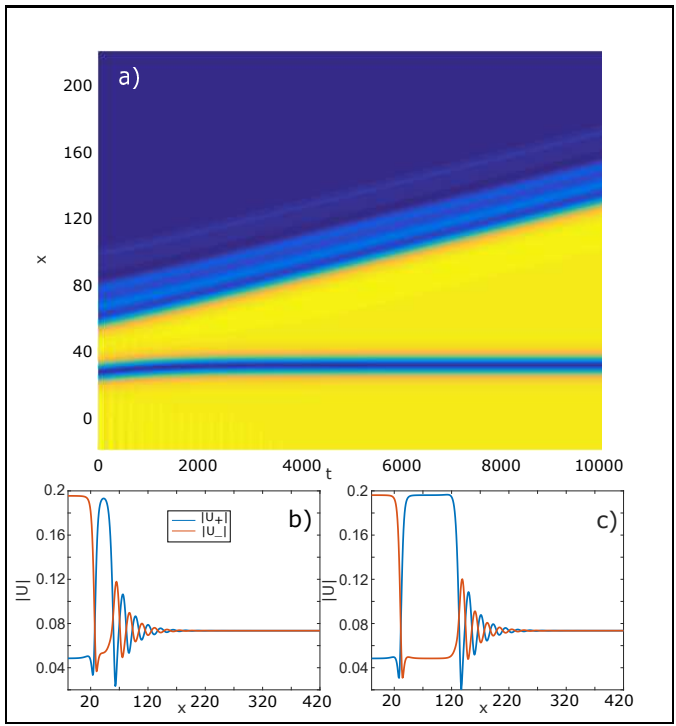}}
\caption{(a) The evolution of the unstable $H_-B$ domain wall at $P = 0.22$ obtained by numerical simulations of equations (\ref{main_1}). The colors shows the total intensity of the state $|U_{+}|^2+|U_{-}|^2$, the blue color corresponds to the lowest intensity and the yellow color - to the higher intensity. (b) Numerically found unstable domain wall taken as the initial conditions for the numerical simulations is shown in (a). The domain wall can be seen as an unstable bound state of $H_+B$ wall and a dark soliton ($H_-BH_+$ soliton) used as an initial condition in the numerical simulation. (c) Field intensity distributions at $t = 10000$.}
\label{fig5}
\end{figure}

Another kind of domain walls considered in this paper is $H_{-}B$ and related to them by the inversion of $x$ axis and the swap of the fields $BH_{+}$ states. Their bifurcations diagrams are shown in Fig.~\ref{fig2}(b) by the red and the yellow curves. The typical field distributions in the $BH_{+}$ for different pump intensities are shown in Fig.~\ref{fig3}(g)-(i). At low pumps the domain walls are wide with the width going to infinity at the left end of the bifurcation diagram. 

At some pump the domain wall experience a fold bifurcation and become unstable. The domain wall belonging to the lower unstable branch of the bifurcation diagram can be seen as an equilibrium combination of a $BH_{+}$ domain wall and a dark soliton. This becomes obvious at the left end of the bifurcation curve where the distance between the domain wall and the soliton goes to infinity. 

The structure of these domain walls suggests the possible mechanism of their instability. If the distance between the soliton and the domain wall becomes smaller then the soliton gets attracted and collide with the domain wall. Alternatively if the distance between the soliton and the domain wall increases then they depart from each other. Numerical simulations confirmed this guess fully. The development of the $H_{-}B$ domain wall instability is illustrated in Fig.~\ref{fig5}. It is clearly seen that the development of the instability results in the formation of a resting dark soliton and the domain wall moving away from the soliton. 

Considering the nature of the instability it can be concluded that the instability growth rate becomes small at low pumps where the domain wall and the dark soliton are well separated and, thus, interact very weakly. At the fold bifurcation the instability growth rate also goes to zero, thus, the maximum instability growth rate is inside the region of the domain wall existence. The numerical analysis shows that the point of the maximal instability growth rate is shifted towards the fold bifurcation point.

In this Section it is shown that in the considered system there are several kinds of domain walls and that some of the domain walls are stable. It was also shown that the instability can result in the formation of dark solitons that can be seen as stable bound states of two domain walls. This calls for a systematic studies of the dissipative solitons that occur in the system. This is the subject of the next Section.

\section{Dark solitons}

In this Section we discuss different dissipative solitons that can be interpreted as bound states of the domain walls connecting the lower bright and the hybrid spatially uniform states. 

We start with the solitons that are formed by the domain wall $H_{+}B$ on the left and domain wall $BH_{-}$ on the right. As it is discussed above, the hybrid states forming these domain walls are related by the operation of simultaneous inversion of spatial coordinate and the swap of the fields $U_{\pm}$. Each of the hybrid states $H_{+}$ on the left and  $H_{-}$ on the right have the energy flow directed towards the bound state formed by the domain walls. The intensity of the field in the centre of the bound state is lower than the intensity of the backgrounds and so the bound state can be refereed as a dark dissipative soliton. We can denote the soliton as $H_{+}BH_{-}$ soliton meaning that the the soliton is the state consisting of the $H_{+}B$ on the left and $BH_{-}$ on the right.

\begin{figure}[t]
\centering
\fbox{\includegraphics[width=\linewidth]{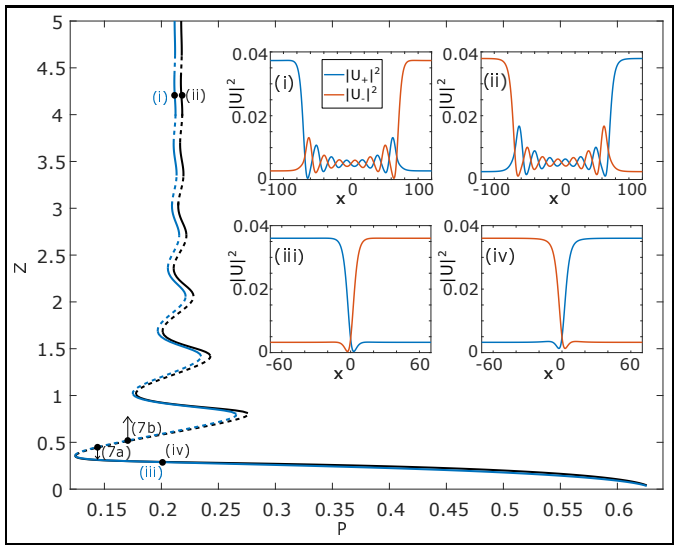}}
\caption{ Bifurcation diagrams of  $H_+BH_-$ (blue line) and $H_-BH_+$ (black line) solitons. The dashed lines correspond to the unstable states. It is seen that in the vicinity of the Maxwell points the bifurcation curves form snaking patterns. 
Insets (i) and (ii) illustrate the field distributions of the solitons belonging to the snaking patterns of the bifurcation diagrams. These states corresponds to the point (i) and (ii) of the bifurcation curves. The field distributions in the $H_+BH_-$ and $H_-BH_+$ solitons at pump $P = 0.2$ are presented in insets (iii) and (iv) for $H_+BH_-$ and $H_-BH_+$ solitons correspondingly. The relevant points on the bifurcation curves are marked as (iii) and (iv) in main panel. }
\label{fig6}
\end{figure}

We calculated the bifurcation diagram of the soliton, the curve showing the dependence of the soliton width defined as $Z = \int_{-\infty}^{\infty} ||U_+|^2 + |U_-|^2 - (|U_+^0|^2 + |U_-^0|^2)| dx$ on the pump amplitude $P$ is shown in main panel of Fig.~\ref{fig6} by the blue line. Strictly speaking, $Z$-value is not really width, but assuming soliton depth near the Maxwell point to be constant, $Z$ depends only on soliton width. The solitons exist in the whole domain of the existence of the $H$ states. As it is expected, because of the symmetry of the domain walls,  the discussed solitons have zero velocity. 

The linear analysis of the stationary linear perturbations shows that they decay to the bright state $B$ with oscillations and this gives a reason to expect that there may exist more then one equilibrium distance between the domain walls forming $H_{+}BH_{-}$ solitons. It is also important that $H_{+}B$ and $BH_{-}$ domain walls have Maxwell points and that these Maxwell points are at the same pump value for both of the domain walls. To form a bound state of the resting domain walls the interaction strength can be arbitrary weak. Therefore one can expect the formation of the snaking pattern of the bifurcation characteristic around the Maxwell point \cite{snaking1,snaking2,snaking3,bound_st1}. The numerical simulations confirmed that the snaking takes place in the considered case too, see Fig.~\ref{fig6} where the snaking pattern is clearly seen. 

After each of the turn of the snaking pattern the width of the dissipative soliton increases by the period of the oscillations of the domain wall tail decaying to the bright state. The distribution of the filed in a soliton in the vicinity of the Maxwell point is shown in the inset (i) of Fig.~\ref{fig6}. It is seen that the soliton can indeed be considered as two remote and, thus, very weakly interacting domain walls.

\begin{figure}[b]
\centering
\fbox{\includegraphics[width=\linewidth]{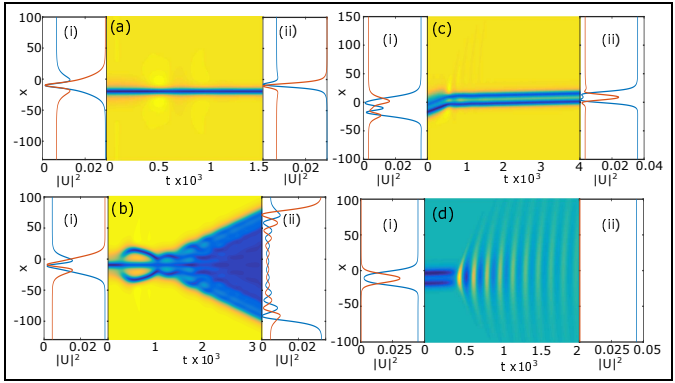}}
\caption{(a) Development  of  the  instability  of $H_+BH_-$ soliton at $P = 0.14$. (i) Dynamically unstable soliton is taken as an initial condition in numerical simulation. (ii) Resulting stable soliton with the width smaller than the initial one. (b) Development  of  the  instability  of $H_+BH_-$ soliton at $P = 0.17$. (i) Initial conditions in the form of an unstable soliton . (ii) The final distributions of the fields where the separation of two domain walls is clearly seen. (c) Development  of  the  instability  of $H_+BH_+$ soliton at $P = 0.2$. (i) The initial condition in the form of the unstable soliton. (ii) Resulting stable soliton with the width smaller than the initial one. (d) Development  of  the  instability  of $H_+BH_+$ soliton at $P = 0.28$. (i) The initial condition in the form of the unstable soliton. (ii) The final spatially uniform field distribution corresponding to hybrid state which appears after the collapse of the soliton. }
\label{fig7}
\end{figure}

The spectral analysis has revealed that in the snaking pattern the stability of the dark solitons $H_{+}BH_{-}$ changes at each fold bifurcation and so the stable parts of the bifurcation curve interchanges with the unstable ones. The instability can be understood in terms of the effective potential created by a domain wall for its neighbor. The maxima of the potential correspond to the unstable solitons. Then one can anticipate that the development of the instability leads to the change of the distance between the domain walls. The results of the instability is the formation of a stable bound state with the width smaller or larger than the width of the initial bound state. The results of the numerical simulations illustrating the development of the instability of  $H_{+}BH_{-}$ solitons are shown in Fig.\ref{fig7}(a). One can see that, indeed, the instability simply changes the width of the soliton to the width of a stable soliton. Otherwise, the instability of the bound state can result in separation of the $H_{+}B$ and $BH_{-}$ domain walls, as it is shown in Fig. \ref{fig7}(b).

Outside the snaking pattern the solitons exist in the whole region of existence of the domain walls. The width of the soliton can be of the size or more narrow then the size of the domains wall, see inset (iii) of Fig.~\ref{fig6}.  The stability analysis tells that the outside the snaking pattern solitons $H_{+}BH_{-}$ are stable provided that the $H_{+}$ and $H_{-}$ states are stable.

Another possible combination of the domain walls that can be in equilibrium and, thus, can be considered as a dissipative soliton is $H_{-}B$ and $BH_{+}$ domain walls. These domain walls form the bound states where the energy on the left and on the right flows away from the bound state. The intensity distribution in these solitons is symmetric and the solitons are at rest, see insets (ii) and (iv) in Fig.~\ref{fig6}. 

In $H_{-}BH_{+}$ solitons the energy flows away from the solitons and, of course, these solitons cannot be identical to $H_{+}BH_{-}$ solitons, however their properties are similar. In particular, the bifurtacion diagram of  $H_{-}BH_{+}$ solitons shown in Fig.~\ref{fig6} by the black curve has a snaking pattern around the Maxwell point. As in the case of  $H_{+}BH_{-}$ solitons, each turn of the bifurcation curve corresponds to the increase of the soliton width by an oscillation period of the domain wall tails decaying to the BS. In the snaking pattern the stability of the soliton changes at each turn of the bifurcation characteristics. The mechanism of the instability is also the same as in the case of  $H_{+}BH_{-}$ solitons: the solitons changing their size to the size of the neighbouring stable soliton. The  $H_{-}BH_{+}$ solitons exist in the range of the existence of $H_{-}B$ and $BH_{+}$ domain walls. Outside the snaking area the solitons are stable provided that the HS on the left and on the right of the soliton  are stable.

\begin{figure}[t]
\centering
\fbox{\includegraphics[width=\linewidth]{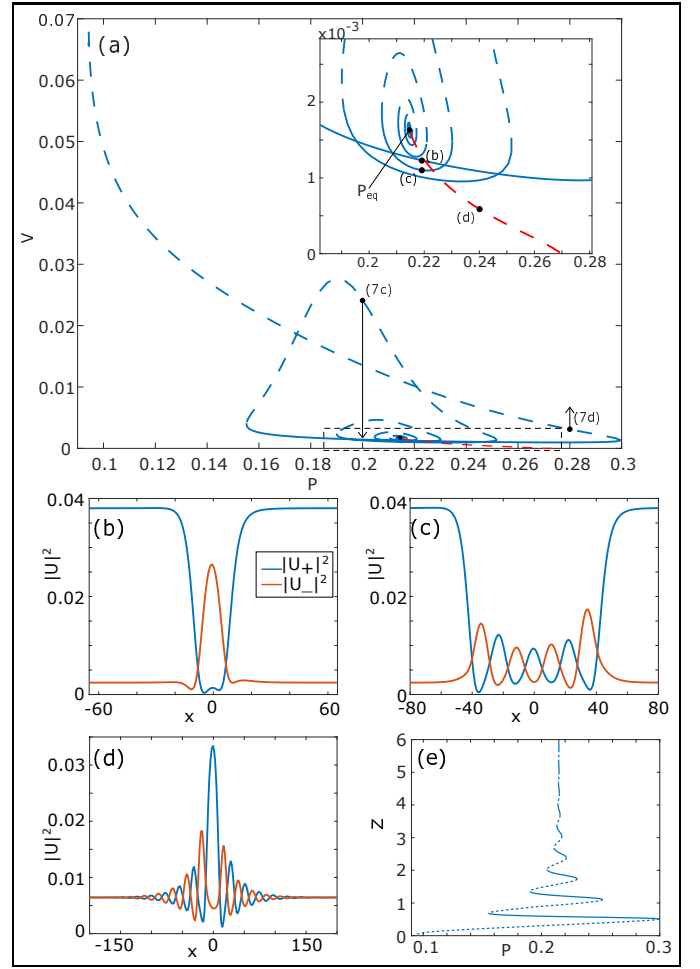}}
\caption{(a) The velocities of $H_{+}BH_{+}$ (blue line) and  $BH_{+}B$ (red line) moving solitons as function fo the pump $P$.  ``$P_{eq}$'' is the point, where the domain walls $H_+B$ and $BH_+$ have the same velocities, see the inset in panel (b) of Fig.~\ref{fig2}. The fields distributions of single-hump $H_+BH_+$ soliton at $P = 0.22$, multi-hump $H_+BH_+$ soliton at $P = 0.22$ and $BH_+B$ soliton at $P=0.24$ are shown in panels (b)-(d) correspondingly. The snaking of the bifurcation curve of the moving  $H_{+}BH_{+}$ soliton around the Maxwell point is shown in panel (b) in the $Z$-$P$ axis.}
\label{fig8}
\end{figure}

A different example of dissipative solitons is the bound states $H_{+} B H_{+}$ and $H_{-} B H_{-}$. The latter ones can be obtained from the former ones by the inversion of $x$ axis and the swap of the fields $U_{\pm} \rightarrow U_{\mp}$. Therefore they possess the same properties and so we discuss here only $H_{+} B H_{+}$ solitons. The peculiarity of these solitons is that because of the symmetry breaking bifurcation the energy flows to the soliton from the left and flows out of the soliton on the right. Thus, one can expect that these solitons are the moving ones. We found numerically the solitons $H_{+} B H_{+}$ and found out that, indeed, the solitons move with some velocity depending on the intensity of the pump, see the bifurcation diagram shown in Fig.~\ref{fig8}(a) by the blue curve. The distribution of the field in the solitons are illustrated in panels (b) and (c) for different points of the bifurcation curve.

As it is seen in Fig.~\ref{fig2}(b) there is a pump $P_{eq}$ at which the domain walls  $H_{+} B$ and  $BH_{+}$ move with the same velocity $v_{eq}$. One can expect that at this point the bound state can be infinitely wide consisting of two not interacting domain walls. Since the stationary fields decay to the BS with oscillations there may be more than one equilibrium distance between the domain walls forming the soliton. At the pump $P_{eq}$ the number of the solitons goes to infinity and, thus, the bifurcation curve swirls toward the point $P=P_{eq}$, $v=v_{eq}$ on the bifurcation diagram plotted in $v-P$ axes. In the bifurcation diagram in the axes $Z-P$ this results in the formation of the snaking pattern discussed above, see Fig.~\ref{fig8}(e).

The problem of the stability of the solitons is of importance from the physical point of view. We studied the stability of the $H_{+} B H_{+}$ solitons by finding the eigenvalues governing the dynamics of small perturbation imposed on the soliton. The analysis shows that $H_{+} B H_{+}$ can be dynamically stable, see Fig.~\ref{fig8}(a) where the solid parts of the bifurcation curve correspond to stable dissipative solitons. The stability of the solitons has been also checked by direct numerical simulations of the partial differential equation with the initial conditions taken in the form of the soliton perturbed by a weak noise. The simulations confirmed the prediction of the spectral stability analysis and shed light on possible outcomes of the instability of  $H_{+} B H_{+}$ solitons. It turned out that the soliton can change its width to that corresponding to a stable soliton, see Fig.\ref{fig7}(c). Alternatively, the bound state of the domain walls forming an unstable bound state can be broken and then domain walls annihilate and the system switches to spatially uniform $H_{+}$ state, as it is shown in Fig.\ref{fig7}(d).

Let us make here a remark that the bound states of the domain walls can also be bright solitons $BH_{\pm}B$ with the intensity having maximum within the soliton. The bifurcation curve of these solitons is shown by red dashed line in  Fig.~\ref{fig8}(a), a typical distribution of the fields of the soliton is illustrated in panel (d). However, we did not find stable solitons of such a kind. This make them less interesting from the physical point of view and we do not discuss them in detail. We note that stable bright solitons exist in the system \cite{OL_BIC_solitons} but these solitons can hardly be considered as bound states of dissipative domain walls but rather as a generalization of bright Nonlinear Schrödinger Equation solitons for the case of driven-dissipative systems. 

\begin{figure}[t]
\centering
\fbox{\includegraphics[width=\linewidth]{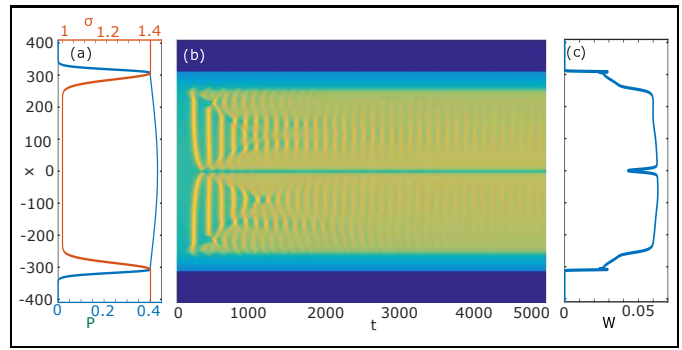}}
\caption{The formation of the $H_-BH_+$ dark soliton from the weak noise taken as the initial conditions. The distributions of the amplitude of the spatially nonuniform pump $P$ is  shown in panel (a) by the blue curve. The distribution of the coupling strength coefficient $\sigma$ is shown in this panel by the red curve. Panel (b) shows the numerically obtained evolution of the filed. The final distribution of the field is shown in panel (c) where the dark soliton is clearly seen. }
\label{fig9}
\end{figure}

The discussed solitons are stable and, thus, can be observed in real experiments. However, the question how these states can be prepared is of great importance from the experimental point of view. At the end of this section we suggest the protocols that allow to form the dissipative solitons starting from the initial condition in the form of weak noise. First we do it for the solitons $H_{-} B H_{+}$. 
To observe the solitons in the numerical simulations we consider a linear system of finite length driven by an external pump whose intensity and distribution in space can be controlled. To avoid strong influence of the boundaries we take the interaction strength between the counter propagating waves to be higher in the vicinity of the edges, in the middle of the system the interaction strength is a constant. The profile of the coupling strength is used in our simulations is shown by red line in panel (a) of Fig.~\ref{fig9}. Experimentally this can be achieved by the increase of the modulation depth of the waveguide at its edges.  

When the pump is switched on the intensities of the field start growing and if the pump is strong enough then at some moment the hybrid states form. However, the boundaries introduces additional losses and, thus, the hybrid state forming at an edge has energy flow directed towards the edge. As a result of the evolution the final state consists of two hybrid states with energy flows directed towards the edge nearest to the state. In the centre of the system the states are connected by a dark soliton, see Fig.~\ref{fig9}(b). The profile of the field intensity of the stationary solution is shown in panel (c) of the Fig. \ref{fig9} and it was checked that this field distribution coincides with the field of the $H_{-}BH_{+}$ soliton found for the given intensity of the pump. 

\begin{figure}[thb]
\centering
\fbox{\includegraphics[width=\linewidth]{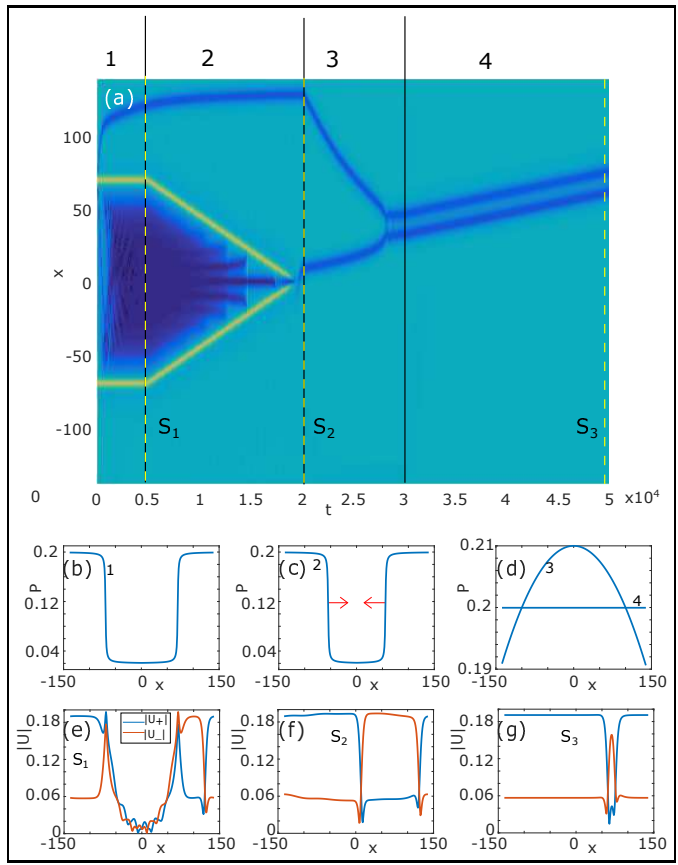}}
\caption{Numerical simulation supporting the protocol of the formation of moving $H_+BH_+$ solitons suggested in the paper. Panel (a) shows the temporal evolution of the intensity of the field. The intervals of different pumping regimes are marked by "$1$"-"$4$".  The field distributions at the times marked by the vertical lines $S_{1-3}$ are presented in panels (e)-(g) correspondingly. The profile of the pump acting on the system within the interval "$1$" ($t < 5000$) is shown in panel (b). The quasi-stationary field forming by $t=5000$ is shown in panel (e). The distribution of the pump amplitude within the time interval "$2$" ($5000 < t < 20000$) is shown in panel (c). The area of the low intensity of the pump is slowly shrinking and finally the pump becomes spatially uniform. Schematically the shrinking of the low pump area is indicated by the red arrows. The field distributions at $t=20000$ is illustrated in panel (f). The pump distributions shown in panel (d) are for time intervals "$3$" ($20000 < t < 30000$) and "$4$" ($t > 30000$), they are marked by the numbers "$3$" and "$4$" correspondingly. The final distribution of the field ($t=50000$) is shown in panel (g). This field distribution perfectly coinside with the field distribution in the $H_+BH_+$ soliton existing at the pump $P=0.21$.}
\label{fig10}
\end{figure}

Let us mention that to reduce the time of the formation of the state the pump is taken to be slightly inhomogeneous (it is shown in Fig.~\ref{fig9}(a) by blue line). Otherwise, the soliton  forms at a random point in the central part of the system and then slowly drifts towards the centre. This motion happens because of the interaction of the soliton with the boundaries which is very weak for long systems. This makes the soliton motion to be very slow and drastically increases the simulation time needed to observe the formation of a stationary state.

To observe the formation of moving solitons of $H_{\pm}BH_{\pm}$ it is more convenient to use an annular system with periodic boundary conditions. First, we need to grow a uniform hybrid state from initial weak noise. To do it, the uniform pump with intensity sufficient to set in the instability of the bright state is used. The development of the instability creates a stationary spatially uniform hybrid state with nonzero energy flow. The hybrid state forming at the pump $P = 0.2$ is the initial conditions for the protocol of the creation of the moving dissipative solitons. To obtain a non-uniform field distribution we change the pump to the spatially nonuniform one shown in Fig.~\ref{fig10}(b). In the area of low intensity the pump cannot support the hybrid state and a stationary low-intensity bright state form in this region. Thus, in the system appears the areas of the contact of the bright and the hybrid states. 

It is possible to choose the difference of the minimum and the maximum pump intensities so that the hybrid states forming in the vicinity of the contact areas have the energy flow directed toward the bright state (to the area of the lower pump). This means that the interval filled with the bright state is between  different hybrid states (one $H_{+}$ and another $H_{-}$). But the system is annular and, thus, there must be an area were  $H_{+}$ and $H_{-}$ states contact each other. This results in the formation of a dark dissipative soliton, see Fig.~\ref{fig10}(a) where in the region marked as ``$1$'' ($0<t<5000$) this dark soliton is clearly seen as well as two areas of contact of the bright and hybrid states. The distribution of the quasi-stationary fields as functions of the $x$ coordinate are also shown in Fig.~\ref{fig10}(e).

Then, at $t=5000$ we start changing the pump gradually reducing the width of the low intensity pump, see Fig.~\ref{fig10}(c). At $t=20000$ the pump becomes spatially uniform and a new dark dissipative soliton $H_{+}BH_{-}$ appears in the system. The region marked as "$2$" in  Fig.~\ref{fig10}(a) shows this process. So now there are two dark solitons in the system, one of them is  $H_{+}BH_{-}$ soliton and another is $H_{-}BH_{+}$ soliton, see Fig.~\ref{fig10}(f) showing the distribution of the fields.

A moving $H_{+}BH_{+}$ or $H_{-}BH_{-}$ soliton can now be created by a gentle collision of the two dark solitons. To do so we slightly change the shape of the pump making it slightly nonuniform. The dark solitons $H_{+}BH_{-}$ and  $H_{-}BH_{+}$ move in opposite directions in the nonuniform pump, one soliton moves down and the other moves against the gradient of the pump intensity. This process is seen in the time interval $20000<t<30000$ marked as "$3$" in  Fig.~\ref{fig10}(a) for the pump shown in Fig.~\ref{fig10}(d). After collisions of the solitons a moving soliton $H_{+}BH_{+}$ appears, the fields distribution in this soliton is illustrated in Fig.~\ref{fig10}(g). It is worth noting here that changing the shape of the pump it is possible to swap the sides of the colliding solitons and, thus, produce a  $H_{-}BH_{-}$ soliton.


At $t=30000$ the pump is made uniform again and one can see that there is a stable dissipative soliton moving  in the system (region "$4$" in Fig.~\ref{fig10}(a)). Thus, we can conclude that the suggested protocol allows to create the dissipative solitons moving either clock or counter-clockwise in the pumped annular system.   Let us remark here that the suggested technique can also be used for the creation of more complex soliton structures.

So we can say that the protocols of possible experimental observation of the solitons are suggested and verified. This brings us to the end of this section and we proceed to the next one where the main results of the work are summarized.

\section{Conclusion}

In this paper we have considered the switching waves connecting the states with broken (hybrid states) and unbroken symmetries (bright states). As it is discussed in the main text the states with broken symmetry have nonzero energy flow and, thus, the domain walls depends on the relative positions of the connected states (the domain wall connecting the BS on the left and HS on the right is not equivalent to the domain wall connecting the HS on the left to the BS on the right). The domain walls are classified and their bifurcation diagrams are found and discussed. 

It is shown that the domain walls can form a bound states that can be called dissipative solitons. An interesting finding is that the disipative solitons can be moving and the direction of the motion is defined by the symmetries of the soliton background. The stability of the dissipative solitons is studied and it is shown that the solitons can be stable and so can be observed experimentally. In the paper we also suggested and verified numerically the protocols allowing to create the discussed solitons.

\begin{acknowledgments}
The authors acknowledge the financial support provided by Russian Fund for Basic Research (Grant ``Aspiranty'' No. 20-32-90227) and by the  Ministry of Science and Higher Education of the Russian Federation (Megagrant number 14.Y26.31.0015). 
\end{acknowledgments}

\bibliography{sample}

\end{document}